\documentclass[runningheads]{llncs}
\usepackage[T1]{fontenc}
\usepackage[misc]{ifsym}
\usepackage{graphicx}
\usepackage{amsmath}
\usepackage{multirow} 
\usepackage{booktabs}
\usepackage{url}
\begin{document}
\title{FLeW: Facet-Level and Adaptive Weighted Representation Learning of Scientific Documents}
\titlerunning{Facet-Level and Adaptive Weighted Representation Learning}
%
\author{
    Zheng Dou\textsuperscript{1} \and
    Deqing Wang\textsuperscript{1,3(\Letter)} \and
    Fuzhen Zhuang\textsuperscript{2,3} \and
    Jian Ren\textsuperscript{1} \and
    Yanlin Hu\textsuperscript{4}
}
%
\authorrunning{Z. Dou et al.}
%
\institute{
    SKLSDE, School of Computer Science, Beihang University, Beijing, China \\ \email{\{miracle\_dz, dqwang, renjian\}@buaa.edu.cn}\and
    Institute of Artificial Intelligence, Beihang University, Beijing, China \email{zhuangfuzhen@buaa.edu.cn}\and
    Zhongguancun Laboratory, Beijing, China \and
    National Computer Network Emergency Response Technical Team/Coordination Center of China, Beijing, China \\ \email{yanlinhu@cert.org.cn}
    %
}
\maketitle              
\begin{abstract}
    Scientific document representation learning provides powerful embeddings for various tasks, while current methods face challenges across three approaches. 1) Contrastive training with citation-structural signals underutilizes citation information and still generates single-vector representations. 2) Fine-grained representation learning, which generates multiple vectors at the sentence or aspect level, requires costly integration and lacks domain generalization. 3) Task-aware learning depends on manually predefined task categorization, overlooking nuanced task distinctions and requiring extra training data for task-specific modules. To address these problems, we propose a new method that unifies the three approaches for better representations, namely FLeW. Specifically, we introduce a novel triplet sampling method that leverages citation intent and frequency to enhance citation-structural signals for training. Citation intents (background, method, result), aligned with the general structure of scientific writing, facilitate a domain-generalized facet partition for fine-grained representation learning. Then, we adopt a simple weight search to adaptively integrate three facet-level embeddings into a task-specific document embedding without task-aware fine-tuning. Experiments show the applicability and robustness of FLeW across multiple scientific tasks and fields, compared to prior models.

\end{abstract}
\section{Introduction}
The rapid growth of scientific publications across diverse fields has spurred the need for high-quality document representations to support downstream tasks like classification, retrieval, and search \cite{specter,scirepeval,ban2025pagerank}. Compared to general-purpose text, scientific documents exhibit unique relational structures and encapsulate more concentrated knowledge. Current approaches focus on these features of scientific documents for better representations while still facing challenges.

First, contrastive training with citation-structural signals has shown effectiveness in scientific document representation learning \cite{specter,scincl22}. However, they underutilize richer information from citation edges and generate single-vector representations, limiting their ability to capture fine-grained information. Second, fine-grained representation learning generates multiple vectors at the sentence or aspect level to capture detailed knowledge and information \cite{aspire,specialized}. However, these methods either require costly integration or lack generalization across different fields. Third, task-aware learning also generates multi-vector representations, with each vector tailored to a specific task category \cite{scimult,scirepeval}. But they need manually predefined task categorization and fail to capture nuanced differences within the same category. Additionally, such methods often require extra modules and additional training data to learn task-specific parameters.

In this work, we propose \textbf{\underline{F}}acet-\textbf{\underline{L}}evel and Adaptiv\textbf{\underline{e}} \textbf{\underline{W}}eighted representation learning of scientific documents, named \textbf{FLeW}, which unifies citation-structural contrastive training, fine-grained multi-vector representation, and task-aware learning into a single framework and address their challenges. Specifically, FLeW uses a novel triplet sampling method to utilize citation intent (background, method, result) and citation frequency for more informative contrastive training. The classification of citation intent aligns with the general structure of scientific writing, serving as a generalized facet partition for fine-grained multi-vector learning with more applicability across different fields. FLeW adopts a simple weight search to use the weighted sum of facet-level representations as the final document representation adaptive to multiple tasks instead of task-aware fine-tuning. By integrating task information into facet weights, FLeW effectively captures task-specific differences and improves the applicability across tasks. In summary, the contributions of this work are as follows: 
\begin{itemize}
    \item We propose a structural sampling method that uses citation intent and frequency information for enhanced citation-structural learning, along with a textual splitting method to divide triplet texts into faceted parts for better fine-grained textual representation. 
    \item We propose to adopt a simple weight search and use the weighted sum of facet-level representations as the final document representation adaptive to multiple tasks instead of task-aware fine-tuning. 
    \item Extensive experiments on a large-scale benchmark comprising 19 tasks and a citation recommendation dataset spanning 19 fields demonstrate that FLeW achieves superior performance compared to prior methods across multiple scientific tasks and fields.
\end{itemize}

\section{Background}
Document representation learning is the process of encoding a given input document $\mathcal{D}$ into a low-dimensional dense vector $\boldsymbol{V}$ by an ${Encoder}$ model. The generated vector, also known as embedding or representation, contains rich semantic and context information of the document and can be used for vector calculation in downstream tasks. BERT \cite{bert} is a pre-trained model with encoder-only architecture consisting of multiple layers of Transformer \cite{transformer} to encode the tokens in a given input sequence. The final hidden state of special ${[CLS]}$ token is commonly used as an aggregate representation of the input sequence. 

For a scientific document, the title provides a thorough overview of the paper's topic and the abstract is a comprehensive summary of the entire text. Following prior work \cite{specter,scincl22,scirepeval}, using the ${title}$ and ${abstract}$ separated by the ${[SEP]}$ token as input effectively balances input length constraints with high-quality representation. 
Since the great success of BERT on various NLP tasks especially text understanding, we use BERT-based model as the ${Encoder}$ and also take the final hidden state of the ${[CLS]}$ token as output representation:
\begin{equation}
\label{eq1}
\begin{aligned}
    \boldsymbol{V} = Encoder(title[SEP]abstract)_{[CLS]} 
\end{aligned}
\end{equation}
\indent Our goal is to obtain such pre-trained ${Encoder(s)}$ which can generate informative representation of any input scientific document (title and abstract) directly utilized for downstream tasks.
\begin{figure}[!b]
    \centering
    \includegraphics[width=1\textwidth]{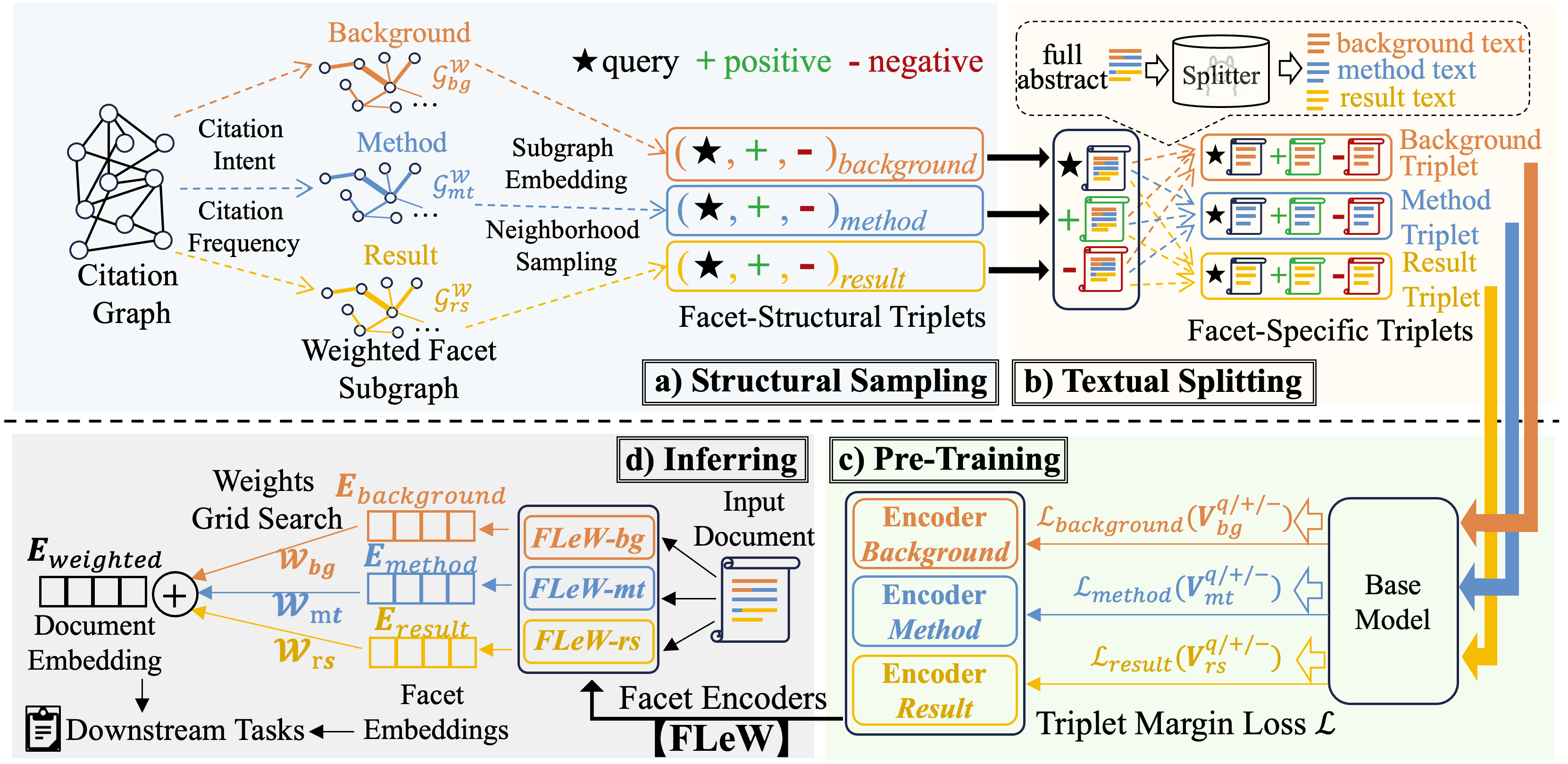} 
    \caption{Approach overview of FLeW.}
    \label{fig1}
\end{figure}

\section{Methodology}

\subsection{Structural Sampling}
A citation from a citing paper consists of one or several sentences referring to a specific aspect of the cited paper. This aspect, commonly referred to as citation intent, can be classified into three distinct categories: background, method, and result. Furthermore, the influence of cited papers within the same citing paper can vary significantly. An intuitive criterion is that the more frequently a cited paper is referenced within the citation contexts of a citing paper, the greater its influence on the citing paper. Hence, citation edges can both be classified into three intents and be influence-weighted based on citation frequency. As shown in Fig.\ref{fig1}(a), we propose a novel triplet sampling method utilizing this information to sample facet-structural triplets from citation graph for enhanced citation-structural contrastive learning. 

\subsubsection{Weighted Facet Subgraphs.}  
We construct the original citation graph $\mathcal{G} = (\mathcal{V}, \mathcal{E})$, where $\mathcal{V}$ represents the set of paper nodes, and $\mathcal{E}$ denotes the set of citation edges between citing and cited papers. Each edge $e \in \mathcal{E}$ is  associated with citation contexts and annotated with citation intent labels ($\mathcal{T} = \{background\,(bg), method\,(mt), result\,(rs)\}$). Based on the intent labels, we extract three facet-specific subgraphs, $\mathcal{G}_{bg}, \mathcal{G}_{mt}, \mathcal{G}_{rs}$, where each subgraph captures citation edges corresponding to a specific intent. Formally, the facet subgraphs are defined as follows:
\begin{equation}
    \mathcal{G}_{x} = (\mathcal{V}, \mathcal{E}_{x}), \quad \mathcal{E}_{x} = \{e \in \mathcal{E} \mid e^{\mathcal{T} = x}\}, \quad x \in \{bg, mt, rs\}.
\end{equation}

To further incorporate the influence of citation frequency, we assign weights to the edges in each facet subgraph. According to our proposed criterion of influence, for the same citing paper, the cited paper with more citations is more influential than those with less citations. The weight $w(e)$ of an edge $e \in \mathcal{E}_x$ is determined by the number of citation contexts $\vert\mathcal{C}_x(e)\vert$ associated with the same intent $x$ between citing and cited nodes. The resulting weighted facet subgraphs are formalized as:
\begin{equation}
    \mathcal{G}_{x}^W = (\mathcal{V}, \mathcal{E}_{x}, \mathcal{W}_{x}), \quad \mathcal{W}_{x}(e) = \vert\mathcal{C}_x(e)\vert, \quad x \in \{bg, mt, rs\}.
\end{equation}

\subsubsection{Subgraph Embedding and Neighborhood Sampling.} We utilize PyTorch-BigGraph (PBG) \cite{PBG} to train and generate subgraph embeddings in the latent space derived from the citation-structural information. As PBG cannot be directly used to embed weighted graphs, we convert the weighted edges into repeated edges with a frequency equal to their weights. With the structural embedding results of papers in each facet subgraph respectively, distances between paper nodes become measurable. By conducting neighborhood sampling to obtain positive and negative papers based on queries, we finally get facet-structural triplets: 
$(query,positive,negative)_{background/method/result}$.

\subsection{Textual Splitting} 
Our proposed structural sampling not only leverages the intent and frequency information on citation edges, but also samples three sets of facet-structural triplets from three weighted subgraphs as training data. Three types of triplets enable fine-grained multi-vector representations of scientific documents. The partition into three facets (background, method, and result) also aligns with the general structure of scientific writing, enhancing its applicability across fields. As shown by Eq.(\ref{eq1}), abstract texts serve as the primary content for representing scientific documents. But full abstract texts in our facet-structural triplets contain information from all three facets. To further learn facet-specific textual information, we propose splitting the full text of triplet abstracts into three faceted parts and retaining only the part corresponding to the type of triplet facet. The process of textual splitting is illustrated in Fig.\ref{fig1}(b), where we use different colors to represent facet-specific texts.

Prior works \cite{hierarchical,sequential} have defined this task as \textit{Sequential Sentence Classification} and train discriminative models on domain-specific labeled data. Considering multiple fields of our triplet papers and the issue of sentence boundary disambiguation in discriminative models, we propose to finish this task by a generative large language model (LLM) with instruction tuning as follows:
\begin{quote}
\small{
\noindent\textbf{Instruction:} \textit{Here is the abstract of a scientific paper. Your task is to split the text into three distinct sections based on the content of sentences:}

\textit{1.Background: The context or previous knowledge related to the topic.}

\textit{2.Method: The methodology, approach, or contribution proposed in the paper.}

\textit{3.Result: The findings, outcomes, or conclusions derived from experiments or analysis.}

\textit{Please return the output in a structured JSON format, as shown below:} 

$\qquad\{``background": ``xxx",``method": ``xxx",``result": ``xxx"\}$

\textit{Ensure that the original text remains intact in each section and that every sentence is categorized appropriately.}

\noindent\textbf{Input:} $``[abstract]"$

\noindent\textbf{Output:}\{\textit{``background'':``abstract{\scriptsize[bg]}'',``method'':``abstract{\scriptsize[mt]}'',``result'':``abstract{\scriptsize[rs]}''}\}
}
\end{quote}

Following the template above, we first use GPT-4o to generate outputs of 12k papers as training data, and then conduct an instruction tuning on the Llama-3.1-8B-Instruct model \cite{llama} with LoRA \cite{lora} to adapt it to current task as $Splitter$. Finally, we get facet-specific triplets $(query_{[x]}, positive_{[x]}, negative_{[x]})_x$ enriched with both structural and textual information:
\begin{equation}
\begin{aligned}
    (query, pos&itive, negative)_x \xrightarrow{Splitter} 
    (query_{[x]}, positive_{[x]}, negative_{[x]})_x\:,\\ &x \in \{background\,(bg), method\,(mt), result\,(rs)\} 
\end{aligned}
\end{equation}

\subsection{Pre-Training}
With facet-specific triplets from Structural Sampling and Textual Splitting, three encoders for representing facet-specific information are pre-trained on corresponding triplets respectively as shown in Fig.\ref{fig1}(c). Triplet margin loss is used for contrastive learning to make queries and positives closer while queries and negatives further:
\begin{equation}
    \mathcal{L}_x = \max \{ f(\boldsymbol{V}_x^q, \boldsymbol{V}_x^+) - f(\boldsymbol{V}_x^q, \boldsymbol{V}_x^-) + \delta, 0 \}, \quad x \in \{bg, mt, rs\}. 
\end{equation}
\noindent where $f$ is a L2 norm distance function and $\delta$ is the margin hyperparameter ($\delta=1$ as prior works). $\boldsymbol{V}_x$ is obtained following Eq.(1) except a slight difference that the abstract text used here is facet-specific $abstract_{[x]}$ from textual splitter.

The three facet encoders $\text{Encoder}_x$, optimized from the base model through their respective loss functions $\mathcal{L}_x$, serve as the final pre-trained models directly for scientific document representation, achieving the goal outlined in Section 2. 
\begin{equation}
\begin{aligned}
    \text{Enco}&\text{der}_x = f(\cdot; \theta_x^*), \quad \theta_x^* = \arg\min_{\theta} \mathcal{L}_x(\theta), \quad \theta^{(0)} = \theta_{base}\:, \\ &x \in \{background\,(bg), method\,(mt), result\,(rs)\}
\end{aligned}
\end{equation}

\subsection{Inferring}
Considering the contextual information in the full abstract text helpful to document representation and the extra cost of splitting every input abstract into facet-based parts, we use the title and full abstract text as inputs during inference. Enhanced by the facet-textual $abstract_{[x]}$ during training, facet encoders can focus on representing facet-specific information even with the full abstract text. As shown in Fig.\ref{fig1}(d), three pre-trained encoders, referred to as FLeW-bg, FLeW-mt, and FLeW-rs, generate three facet-level embeddings based on the title and full abstract of an input scientific document:
\begin{equation}
\begin{aligned}
    \boldsymbol{E_{background}}=\text{FLeW-bg}(title[SEP]abstract)_{[CLS]}\\
    \boldsymbol{E_{method}}=\text{FLeW-mt}(title[SEP]abstract)_{[CLS]}\\
    \boldsymbol{E_{result}}=\text{FLeW-rs}(title[SEP]abstract)_{[CLS]}\\
\end{aligned}
\end{equation}

\indent We use a simple weighted sum of three representations as the final document representation, where facet weights are adaptive to different downstream tasks and reflect the importance of each facet. $E_{weighted}$ is the document-level embedding of our proposed FLeW:

\begin{equation}
\begin{aligned}
&\boldsymbol{E_{weighted}}=\mathcal{W}_{bg}*\boldsymbol{E_{background}}+\mathcal{W}_{mt}*\boldsymbol{E_{method}}+\mathcal{W}_{rs}*\boldsymbol{E_{result}}\\
&s.t.\qquad\mathcal{W}_{bg}+\mathcal{W}_{mt}+\mathcal{W}_{rs}=1,\mathcal{W}_{bg}>0,\mathcal{W}_{mt}>0,\mathcal{W}_{rs}>0
\end{aligned}
\end{equation}
\indent Because of the limitation that the weight sum must be 1, there are actually only two free weight parameters ranging from 0 to 1. We use a simple grid search to iterate across all weight combinations and rank by evaluation score for each task on the validation dataset to select optimal weights.

\section{Experiments}
\subsection{Implementation Details}
We use the 2024-04-02 release version of Semantic Scholar Academic Graph (S2AG) \cite{S2AG} as our base corpus. In S2AG, we download \textit{citations} dataset to build citation graph for structural sampling, and download \textit{papers} and \textit{abstracts} datasets to obtain metadata (title and abstract) of papers for textual splitting. Then we collect a query list of 216k papers for facet triplet sampling. By conducting neighborhood sampling across three facet subgraphs,we generate 2.16M triplets with 2.95M, 2.81M and 2.94M metadata for training three facet encoders, respectively. During training, we initialize the model with SciBERT \cite{scibert} and use Adam with weight decay as optimizer with the initial learning rate $\lambda=2^{-5}$, following \cite{specter,scincl22}. Our facet encoders are trained on 2 NVIDIA GeForce RTX 3090 (24G) with batch size 10 for 2 epochs. We release our data and encoder models\footnote{\scriptsize \url{https://huggingface.co/collections/Miracle-dz/flew-67b3074bce03f9a0573cd94d}}.

\subsection{Baseline Models}
We compare our method to the base model SciBERT \cite{scibert} and other prior works with similar triplet sampling and contrastive learning approach: SPECTER \cite{specter}, SciNCL \cite{scincl22} and SPECTER-2 \cite{scirepeval}. Besides, to show the distinctiveness of scientific documents, there is also a comparison to SFR-2\_R\footnote{\scriptsize \url{https://huggingface.co/Salesforce/SFR-Embedding-2_R}}, the leader model (2024-06-14) on MTEB benchmark \cite{mteb}, which represents the performance of general-purpose text embedding models.

\subsection{Experimental Results}
\subsubsection{Multiple Tasks Evaluation.} Table \ref{table1} presents the document-level evaluation results across 19 tasks in four formats on SciRepEval benchmark \cite{scirepeval}. Overall, our model FLeW achieves the best performance in 13 out of 19 tasks, and also obtains the second-best performance in 4 out of the remaining 6 tasks. The average performance across the total benchmark achieves the best result (60.81), surpassing the second-best by +0.60. FLeW also achieves the best score on average across all the four formats.
\begin{table*}[!ht]
    \caption{Document-level evaluation on SciRepEval benchmark across 19 tasks.}
    \label{table1}
    \centering
    \setlength{\tabcolsep}{0.5mm} 
    \renewcommand{\arraystretch}{0.77}
    \small
    \scalebox{0.86}{
    \begin{tabular}{@{}ccccccccccccc@{}}
        \midrule
        Format & \multicolumn{6}{c}{Proximity {[}PRX{]}} & \multicolumn{6}{c}{Regression {[}RGN{]}} \\ \cmidrule(r){2-7} \cmidrule(l){8-13} 
        Task & {\scriptsize Relish} & \begin{tabular}[c]{@{}c@{}}{\scriptsize Revi-}\\ {\scriptsize ewer}\end{tabular} & \begin{tabular}[c]{@{}c@{}}{\scriptsize Same}\\ {\scriptsize Author}\end{tabular} & \begin{tabular}[c]{@{}c@{}}{\scriptsize HighInf}\\ {\scriptsize Citation}\end{tabular} & \begin{tabular}[c]{@{}c@{}}{\scriptsize SciDocs} \\ {\scriptsize PRX}\end{tabular} & \multirow{2}{*}{\begin{tabular}[c]{@{}c@{}}{[}PRX{]}\\ Avg\end{tabular}} & \begin{tabular}[c]{@{}c@{}}{\scriptsize Review}\\ {\scriptsize Score}\end{tabular} & \begin{tabular}[c]{@{}c@{}}{\scriptsize Max} \\ {\scriptsize H-index}\end{tabular} & \begin{tabular}[c]{@{}c@{}}{\scriptsize Tweet}\\ {\scriptsize Mentions}\end{tabular} & \begin{tabular}[c]{@{}c@{}}{\scriptsize Citation}\\ {\scriptsize Count}\end{tabular} & \begin{tabular}[c]{@{}c@{}}{\scriptsize Pub}\\ {\scriptsize Year}\end{tabular} & \multirow{2}{*}{\begin{tabular}[c]{@{}c@{}}{[}RGN{]}\\ Avg\end{tabular}} \\ 
        \cmidrule(lr){2-6} \cmidrule(lr){8-12}        
        Metric & {\scriptsize nDCG} & {\scriptsize Avg. P} & {\scriptsize MAP} & {\scriptsize MAP} & {\scriptsize nDCG} &  & {\scriptsize K Tau} & {\scriptsize K Tau} & {\scriptsize K Tau} & {\scriptsize K Tau} & {\scriptsize K Tau} &  \\ \cmidrule(lr){2-7} \cmidrule(l){8-13}
        {\scriptsize SciBERT} & 82.81 & 41.68 & 79.48 & 33.72 & 76.15 & 62.77 & 20.26 & 12.38 & 22.18 & \underline{39.16} & 27.71 & 24.34 \\
        {\scriptsize SPECTER} & 90.07 & 45.16 & 86.53 & 42.89 & 93.98 & 71.73 & 17.35 & 12.52 & 24.19 & 33.21 & 25.96 & 22.65 \\
        {\scriptsize SciNCL} & 90.67 & 45.40 & \textbf{87.47} & 43.39 & \textbf{95.01} & 72.39 & 18.87 & \underline{13.45} & 25.79 & 34.61 & 28.99 & 24.34 \\
        {\scriptsize SPECTER-2} & \underline{91.63} & \underline{45.42} & 87.00 & \underline{44.96} & \underline{94.87} & \underline{72.78} & \underline{20.63} & 12.76 & \textbf{27.11} & 38.33 & \underline{33.65} & \underline{26.50} \\
        {\scriptsize SFR-2\_R} & 88.36 & 45.56 & 82.09 & 39.52 & 87.10 & 68.53 & 11.18 & 8.31 & 14.09 & 36.24 & 31.23 & 20.21 \\
        {\scriptsize FLeW(ours)} & \textbf{92.15} & \textbf{45.70} & \underline{87.03} & \textbf{45.50} & 94.46 & \textbf{72.97} & \textbf{20.93} & \textbf{15.09} & \underline{27.01} & \textbf{39.21} & \textbf{33.90} & \textbf{27.23} \\ \midrule
        
        Format & \multicolumn{4}{c}{Query {[}QRY{]}} & \multicolumn{7}{c}{Classification {[}CLF{]}} & \multirow{3}{*}{\begin{tabular}[c]{@{}c@{}}Total\\ AVG\end{tabular}} \\ \cmidrule(r){2-5}\cmidrule(lr){6-12}
        Task & \begin{tabular}[c]{@{}c@{}}{\scriptsize NFC-}\\ {\scriptsize orpus}\end{tabular} & \begin{tabular}[c]{@{}c@{}}{\scriptsize TREC}\\ {\scriptsize Covid}\end{tabular} & {\scriptsize Search} & \multirow{2}{*}{\begin{tabular}[c]{@{}c@{}}{[}QRY{]}\\ Avg\end{tabular}} & \begin{tabular}[c]{@{}c@{}}{\scriptsize Biomi-}\\ {\scriptsize micry}\end{tabular} & {\scriptsize DRSM} & {\scriptsize MeSH} & {\scriptsize FoS} & \begin{tabular}[c]{@{}c@{}}{\scriptsize SciDocs}\\ {\scriptsize MAG}\end{tabular} & \begin{tabular}[c]{@{}c@{}}{\scriptsize SciDocs}\\ {\scriptsize MeSH}\end{tabular} & \multirow{2}{*}{\begin{tabular}[c]{@{}c@{}}{[}CLF{]}\\ Avg\end{tabular}} &  \\ \cmidrule(l){2-4} \cmidrule(l){6-11}
        Metric& {\scriptsize nDCG} & {\scriptsize nDCG} & {\scriptsize nDCG} &  & {\scriptsize Wt. F1} & {\scriptsize Wt. F1} & {\scriptsize F1} & {\scriptsize F1} & {\scriptsize F1} & {\scriptsize F1} &  &  \\ \cmidrule(lr){2-5}\cmidrule(lr){6-12} \cmidrule{13-13}
        {\scriptsize SciBERT} & 53.34 & 79.73 & 71.49 & 68.19 & 50.00 & 64.01 & 76.71 & 42.67 & 79.50 & 79.97 & \multicolumn{1}{c|}{65.48} & \multicolumn{1}{c|}{54.37} \\
        {\scriptsize SPECTER} & 64.90 & 86.53 & 73.25 & 74.89 & \textbf{51.22} & \textbf{66.16} & 85.46 & 43.00 & 79.75 & 87.80 & \multicolumn{1}{c|}{68.90} & \multicolumn{1}{c|}{58.42} \\
        {\scriptsize SciNCL} & 70.85 & 87.67 & 73.46 & 77.33 & 50.22 & 65.10 & 86.17 & \underline{43.53} & \underline{81.11} & \underline{89.13} & \multicolumn{1}{c|}{\underline{69.21}} & \multicolumn{1}{c|}{59.52} \\
        {\scriptsize SPECTER-2} & \underline{72.03} & \textbf{89.46} & \underline{73.76} & \underline{78.42} & \underline{50.53} & 62.96 & \underline{86.76} & 42.16 & 81.03 & 89.00 & \multicolumn{1}{c|}{68.74} & \multicolumn{1}{c|}{\underline{60.21}} \\
        {\scriptsize SFR-2\_R} & 67.76 & 86.97 & 70.89 & 75.21 & 47.33 & 63.40 & 81.96 & 38.52 & 78.63 & 86.68 & \multicolumn{1}{c|}{66.09} & \multicolumn{1}{c|}{56.10} \\
        {\scriptsize FLeW(ours)} & \textbf{72.64} & \underline{89.14} & \textbf{74.13} & \textbf{78.64} & 50.25 & \underline{65.41} & \textbf{87.17} & \textbf{44.61} & \textbf{81.71} & \textbf{89.39} & \multicolumn{1}{c|}{\textbf{69.76}} & \multicolumn{1}{c|}{\textbf{60.81}} \\ \midrule
    \end{tabular}
    }
\end{table*}
For each format, FLeW performs best in more than half of the tasks ({\small [PRX]{\scriptsize3/5},[RGN]{\scriptsize4/5},[QRY]{\scriptsize2/3},[CLF]{\scriptsize4/6}}), while the best model for the remaining tasks varies across four formats ({\small [PRX]{\scriptsize SciNCL}, [RGN]\&[QRY]{\scriptsize SPECTER-2}, [CLF]{\scriptsize SPECTER}}). The results show the stability and generalization ability of FLeW across multiple scientific tasks.

\subsubsection{Multiple Fields Evaluation.} Table \ref{table2} presents document-level evaluation results across 19 fields for the citation recommendation task on MDCR dataset \cite{mdcr}. Our model, FLeW, achieves the best performance in almost every field and on average. Compared to other baseline models, SPECTER-2 is more competitive and outperforms our model slightly in several fields. We find that these fields are more humanities-oriented, such as \textit{Philosophy} and \textit{Political Science}, which have fewer ``scientific'' attributes and lack a clear writing structure consisting of three facets (background, method, result). In contrast, FLeW shows significant improvements in more scientific fields which clearly follow the general writing structure, such as \textit{Biology} and \textit{Chemistry}. This demonstrates the effectiveness of our proposed method in capturing the facet-specific information of scientific documents and improving the applicability across multiple scientific fields.

\begin{table*}[!ht]
    \caption{Document-level evaluation on MDCR dataset across 19 fields.}
    \label{table2}
    \centering
    \setlength{\tabcolsep}{0.6mm} 
    \renewcommand{\arraystretch}{0.69}
    \small
    \scalebox{0.88}{
        \begin{tabular}{@{}ccccccccccccc@{}}
            \midrule
            Metric & \multicolumn{6}{c}{MAP} & \multicolumn{6}{c}{Recall@5} \\ \cmidrule(lr){1-1} \cmidrule(r){2-7}\cmidrule(l){8-13} 
            Domain/Model & {\scriptsize BM25} & \begin{tabular}[c]{@{}c@{}}{\scriptsize Sci}\\ {\scriptsize BERT}\end{tabular} & \begin{tabular}[c]{@{}c@{}}{\scriptsize SPEC}\\ {\scriptsize TER}\end{tabular} & \begin{tabular}[c]{@{}c@{}}{\scriptsize Sci}\\ {\scriptsize NCL} \end{tabular} & \begin{tabular}[c]{@{}c@{}}{\scriptsize SPEC}\\ {\scriptsize TER-2}\end{tabular} & \begin{tabular}[c]{@{}c@{}}{\scriptsize FLeW}\\ {\scriptsize (ours)}\end{tabular} & {\scriptsize BM25} & \begin{tabular}[c]{@{}c@{}}{\scriptsize Sci}\\ {\scriptsize BERT}\end{tabular} & \begin{tabular}[c]{@{}c@{}}{\scriptsize SPEC}\\ {\scriptsize TER}\end{tabular} & \begin{tabular}[c]{@{}c@{}}{\scriptsize Sci}\\ {\scriptsize NCL}\end{tabular} & \begin{tabular}[c]{@{}c@{}}{\scriptsize SPEC}\\ {\scriptsize TER-2}\end{tabular} & \begin{tabular}[c]{@{}c@{}}{\scriptsize FLeW}\\ {\scriptsize (ours)}\end{tabular} \\ \cmidrule(lr){1-1} \cmidrule(r){2-7} \cmidrule(l){8-13}
            Art & 38.2 & 22.4 & 34.1 & 34.7 & \underline{43.4} & \textbf{43.7} & 32.3 & 16.6 & 28.8 & 29.2 & \textbf{37.8} & \underline{37.5} \\
            Biology & 38.3 & 20.4 & 34.6 & 36.8 & \underline{39.9} & \textbf{43.5} & 33.6 & 14.0 & 30.0 & 32.3 & \underline{33.3} & \textbf{37.8} \\
            Business & 28.1 & 19.1 & 27.5 & 28.5 & \underline{35.0} & \textbf{35.1} & 22.5 & 13.1 & 21.8 & 24.6 & \textbf{30.5} & \underline{30.0} \\
            Chemistry & 38.0 & 20.0 & 33.7 & 36.5 & \underline{39.7} & \textbf{43.0} & 32.6 & 13.7 & 29.3 & 31.5 & \underline{33.5} & \textbf{38.2} \\
            Computer Science & 34.8 & 19.5 & 35.6 & 37.2 & \underline{38.5} & \textbf{39.9} & 30.5 & 12.8 & 30.4 & 32.2 & \underline{33.4} & \textbf{36.2} \\
            Economics & 30.5 & 21.5 & 27.3 & 28.3 & \underline{33.7} & \textbf{34.7} & 26.0 & 15.4 & 21.9 & 23.2 & \underline{28.5} & \textbf{29.7} \\
            Engineering & 34.6 & 20.5 & 31.3 & 34.2 & \underline{35.4} & \textbf{39.2} & 29.3 & 13.9 & 27.3 & 28.0 & \underline{30.3} & \textbf{33.2} \\
            Environmental Science & 31.6 & 21.3 & 30.1 & 31.5 & \underline{35.0} & \textbf{37.2} & 26.2 & 15.1 & 24.2 & 25.5 & \underline{27.9} & \textbf{32.0} \\
            Geography & 31.8 & 21.9 & 26.4 & 29.5 & \underline{37.1} & \textbf{40.1} & 27.8 & 16.7 & 22.2 & 23.8 & \underline{31.8} & \textbf{35.0} \\
            Geology & 33.1 & 19.5 & 24.8 & 25.7 & \underline{33.4} & \textbf{35.2} & 28.0 & 13.9 & 20.0 & 19.9 & \underline{27.7} & \textbf{29.1} \\
            History & 38.1 & 20.9 & 27.1 & 30.9 & \underline{41.9} & \textbf{43.2} & 32.9 & 15.3 & 20.6 & 23.9 & \underline{34.7} & \textbf{37.6} \\
            Materials Science & 36.1 & 22.2 & 34.1 & 35.8 & \underline{39.7} & \textbf{40.6} & 30.7 & 15.5 & 28.2 & 29.6 & \underline{34.0} & \textbf{35.4} \\
            Mathematics & 35.3 & 22.8 & 34.2 & 34.9 & \underline{40.8} & \textbf{41.2} & 28.3 & 18.3 & 28.9 & 30.1 & \underline{34.2} & \textbf{35.0} \\
            Medicine & 38.6 & 22.0 & 41.4 & 42.7 & \underline{43.8} & \textbf{46.3} & 32.5 & 16.4 & 36.3 & 36.5 & \underline{39.0} & \textbf{41.6} \\
            Philosophy & 30.2 & 19.2 & 27.1 & 29.9 & \textbf{37.2} & \underline{36.6} & 25.7 & 13.3 & 21.1 & 23.5 & \textbf{32.0} & \underline{31.5} \\
            Physics & 35.1 & 23.9 & 30.8 & 34.5 & \underline{37.6} & \textbf{38.4} & 30.2 & 18.2 & 26.3 & 30.3 & \underline{32.5} & \textbf{34.5} \\
            Political Science & 28.6 & 19.4 & 24.2 & 26.4 & \textbf{35.7} & \underline{35.3} & 23.1 & 14.0 & 18.0 & 21.7 & \textbf{31.6} & \underline{30.1} \\
            Psychology & 32.5 & 20.3 & 32.3 & 34.2 & \underline{38.8} & \textbf{40.5} & 28.9 & 16.2 & 28.1 & 30.5 & \underline{33.2} & \textbf{36.1} \\
            Sociology & 26.8 & 20.2 & 25.2 & 26.7 & \underline{34.6} & \textbf{34.8} & 20.5 & 15.8 & 20.5 & 21.9 & \underline{29.8} & \textbf{30.2} \\ \midrule
            Avg & 33.7 & 20.9 & 30.6 & 32.6 & \underline{38.0} & \textbf{39.4} & 28.5 & 15.2 & 25.5 & 27.3 & \underline{32.4} & \textbf{34.2} \\ \midrule
            \end{tabular}
    }
\end{table*}

\subsection{Ablation Study}
Table \ref{table3} presents the results of ablation study on SciRepEval benchmark \cite{scirepeval}. We train three additional facet encoders (FLeW-bg/mt/rs) with the full abstract texts of facet training triplets without textual splitting (\textit{-w/o Textual}) compared to encoders trained with facet-textual triplets (\textit{+w/ Textual}). And SPECTER-2 is used as the representative model from prior works (\textit{-w/o Structural}). 

For facet-level results (FLeW-xx), FLeW-bg outperforms other facet encoders when compared across facets. We attribute this phenomenon to the position bias, as the background facet often appears earlier in the full text, attracting more focus and avoiding the risk of truncation. Our proposed Textual Splitting eliminates this position bias, allowing each facet to be learned equally. The significant improvements of FLeW-rs (\textit{+w/Textual}) provide strong support for this result. 
For document-level results (FLeW), the improvement of performance compared to FLeW-xx indicates the effectiveness of our facet weighted sum strategy, which emphasizes the different importance of each facet adaptive to multiple scientific tasks and fields. Besides, there is a stepwise increase of average performance in each format and total average in the last three rows (e.g. 60.21, 60.66, 60.81). This demonstrates the effectiveness of our proposed Structural Sampling and Textual Splitting, which help to capture the facet-specific information of scientific documents both structurally and textually.

\begin{table*}[!ht]
    \caption{Ablation study on SciRepEval benchmark.}
    \label{table3}
    \centering
    \setlength{\tabcolsep}{0.4mm} 
    \renewcommand{\arraystretch}{0.53}
    \small
    \scalebox{0.83}{
    \begin{tabular}{@{}cccccccccccccc@{}}
        \midrule
        &Format & \multicolumn{6}{c}{Proximity {[}PRX{]}} & \multicolumn{6}{c}{Regression {[}RGN{]}} \\ \cmidrule(r){3-8} \cmidrule(l){9-14} 
        &Task & {\scriptsize Relish} & \begin{tabular}[c]{@{}c@{}}{\scriptsize Revi-}\\ {\scriptsize ewer}\end{tabular} & \begin{tabular}[c]{@{}c@{}}{\scriptsize Same}\\ {\scriptsize Author}\end{tabular} & \begin{tabular}[c]{@{}c@{}}{\scriptsize HighInf}\\ {\scriptsize Citation}\end{tabular} & \begin{tabular}[c]{@{}c@{}}{\scriptsize SciDocs} \\ {\scriptsize PRX}\end{tabular} & \multirow{2}{*}{\begin{tabular}[c]{@{}c@{}}{[}PRX{]}\\ Avg\end{tabular}} & \begin{tabular}[c]{@{}c@{}}{\scriptsize Review}\\ {\scriptsize Score}\end{tabular} & \begin{tabular}[c]{@{}c@{}}{\scriptsize Max} \\ {\scriptsize H-index}\end{tabular} & \begin{tabular}[c]{@{}c@{}}{\scriptsize Tweet}\\ {\scriptsize Mentions}\end{tabular} & \begin{tabular}[c]{@{}c@{}}{\scriptsize Citation}\\ {\scriptsize Count}\end{tabular} & \begin{tabular}[c]{@{}c@{}}{\scriptsize Pub}\\ {\scriptsize Year}\end{tabular} & \multirow{2}{*}{\begin{tabular}[c]{@{}c@{}}{[}RGN{]}\\ Avg\end{tabular}} \\
        \cmidrule(lr){3-7} \cmidrule(lr){9-13}        
        &Metric & {\scriptsize nDCG} & {\scriptsize Avg. P} & {\scriptsize MAP} & {\scriptsize MAP} & {\scriptsize nDCG} &  & {\scriptsize K Tau} & {\scriptsize K Tau} & {\scriptsize K Tau} & {\scriptsize K Tau} & {\scriptsize K Tau} &  \\ \cmidrule(lr){3-8} \cmidrule(l){9-14}
        \multicolumn{1}{c|}{\multirow{3}{*}{\begin{tabular}[c]{@{}c@{}}\it{+ w/} \\ \it{Textual}\end{tabular}}} & {\scriptsize
        FLeW-bg} & \textbf{92.10} & \textbf{45.02} & \textbf{86.54} & \textbf{45.45} & \textbf{93.64} & \textbf{72.55} & \textbf{19.05} & 12.26 & \textbf{26.20} & 36.98 & \textbf{31.36} & \textbf{25.17} \\
        \multicolumn{1}{c|}{}&{\scriptsize FLeW-mt}& \textbf{90.68} & 44.96 & \textbf{86.38} & \textbf{43.96} & \textbf{93.41} & \textbf{71.88} & \textbf{19.19} & 12.95 & \textbf{26.70} & \textbf{38.46} & 32.13 & \textbf{25.89} \\ 
        \multicolumn{1}{c|}{}&{\scriptsize FLeW-rs} & \textbf{89.82} & \textbf{45.14} & \textbf{85.53} & \textbf{42.17} & \textbf{93.30} & \textbf{71.19} & \textbf{20.79} & \textbf{14.83} & 26.13 & \textbf{36.70} & \textbf{32.76} & \textbf{26.24} \\ 
        \cmidrule(l){1-1}
  
        \multicolumn{1}{c|}{\multirow{3}{*}{\begin{tabular}[c]{@{}c@{}}\it{- w/o} \\ \it{Textual} \end{tabular}}}&{\scriptsize FLeW-bg} & 91.99 & 44.79 & 85.99 & 45.41 & 93.21 & 72.28 & 18.91 & \textbf{13.24} & 25.34 & \textbf{37.16} & 30.58 & 25.05 \\
        \multicolumn{1}{c|}{}&{\scriptsize FLeW-mt}& 90.50 & \textbf{45.49} & 86.00 & 43.95 & 93.14 & 71.82 & 18.80 & \textbf{13.82} & 26.17 & 35.11 & \textbf{33.62} & 25.50 \\
        \multicolumn{1}{c|}{}&{\scriptsize FLeW-rs} & 89.47 & 44.86 & 84.66 & 41.57 & 92.71 & 70.65 & 17.27 & 13.82 & \textbf{27.42} & 35.42 & 31.71 & 25.13 \\
        \cmidrule(lr){1-14}
        \multicolumn{2}{c}{{\scriptsize FLeW}} & \textbf{92.15} & 45.70 & \textbf{87.03} & 45.50 & 94.46 & \textbf{72.97} & \textbf{20.93} & \textbf{15.09} & 27.01 & \textbf{39.21} & 33.90 & \textbf{27.23} \\ 
        \multicolumn{2}{c}{\it{- w/o Textual}} & 92.09 & \textbf{45.82} & 86.65 & \textbf{45.60} & 94.22 & 72.88 & 20.23 & 14.78 & \textbf{27.57} & 38.60 & \textbf{34.16} & 27.07 \\
        \multicolumn{2}{c}{\it{- w/o Structural}} & 91.63 & 45.42 & 87.00 & 44.96 & \textbf{94.87} & 72.78 & 20.63 & 12.76 & 27.11 & 38.33 & 33.65 & 26.50 \\
        \midrule
        &Format & \multicolumn{4}{c}{Query {[}QRY{]}} & \multicolumn{7}{c}{Classification {[}CLF{]}} & \multirow{3}{*}{\begin{tabular}[c]{@{}c@{}}Total\\ AVG\end{tabular}} \\ \cmidrule(r){3-6}\cmidrule(lr){7-13}
        &Task & \begin{tabular}[c]{@{}c@{}}{\scriptsize NFC-}\\ {\scriptsize orpus}\end{tabular} & \begin{tabular}[c]{@{}c@{}}{\scriptsize TREC}\\ {\scriptsize Covid}\end{tabular} & {\scriptsize Search} & \multirow{2}{*}{\begin{tabular}[c]{@{}c@{}}{[}QRY{]}\\ Avg\end{tabular}} & \begin{tabular}[c]{@{}c@{}}{\scriptsize Biomi-}\\ {\scriptsize micry}\end{tabular} & {\scriptsize DRSM} & {\scriptsize MeSH} & {\scriptsize FoS} & \begin{tabular}[c]{@{}c@{}}{\scriptsize SciDocs}\\ {\scriptsize MAG}\end{tabular} & \begin{tabular}[c]{@{}c@{}}{\scriptsize SciDocs}\\ {\scriptsize MeSH}\end{tabular} & \multirow{2}{*}{\begin{tabular}[c]{@{}c@{}}{[}CLF{]}\\ Avg\end{tabular}} &  \\ \cmidrule(l){3-5} \cmidrule(l){7-12}
        &Metric& {\scriptsize nDCG} & {\scriptsize nDCG} & {\scriptsize nDCG} &  & {\scriptsize Wt. F1} & {\scriptsize Wt. F1} & {\scriptsize F1} & {\scriptsize F1} & {\scriptsize F1} & {\scriptsize F1} &  &  \\ \cmidrule(lr){3-6}\cmidrule(lr){7-13}\cmidrule{14-14} 

        \multicolumn{1}{c|}{\multirow{3}{*}{\begin{tabular}[c]{@{}c@{}}\it{+ w/} \\ \it{Textual}\end{tabular}}} &{\scriptsize FLeW-bg}& 71.63 & \textbf{88.99} & \textbf{73.69} & \textbf{78.10} & \textbf{48.02} & \textbf{64.14} & \textbf{86.71} & 41.31 & \textbf{81.04} & 87.70 & \multicolumn{1}{c|}{\textbf{68.15}} & \multicolumn{1}{c|}{\textbf{59.57}} \\  
        
        \multicolumn{1}{c|}{}&{\scriptsize FLeW-mt} & 68.06 & \textbf{87.51} & \textbf{73.34} & \textbf{76.30} & \textbf{48.92} & \textbf{64.28} & 85.68 & \textbf{43.35} & \textbf{80.95} & 88.43 & \multicolumn{1}{c|}{\textbf{68.60}} & \multicolumn{1}{c|}{\textbf{59.44}} \\  
        
        \multicolumn{1}{c|}{}&{\scriptsize FLeW-rs} & 67.31 & 85.23 & \textbf{73.02} & 75.19 & \textbf{49.53} & \textbf{62.06} & \textbf{85.39} & 42.39 & \textbf{80.90} & \textbf{88.32} & \multicolumn{1}{c|}{\textbf{68.10}} & \multicolumn{1}{c|}{\textbf{59.02}} \\ 
        \cmidrule(l){1-1}
        \multicolumn{1}{c|}{\multirow{3}{*}{\begin{tabular}[c]{@{}c@{}}\it{- w/o} \\ \it{Textual}\end{tabular}}}&{\scriptsize FLeW-bg}& \textbf{71.65} & 88.64 & 73.68 & 77.99 & 47.82 & 62.44 & 86.19 & \textbf{42.01} & 80.35 & \textbf{88.72} & \multicolumn{1}{c|}{67.92} & \multicolumn{1}{c|}{59.37} \\

        \multicolumn{1}{c|}{}&{\scriptsize FLeW-mt} & \textbf{68.87} & 86.59 & 73.12 & 76.19 & 48.74 & 64.26 & \textbf{86.37} & 41.14 & 80.80 & \textbf{89.01} & \multicolumn{1}{c|}{68.39} & \multicolumn{1}{c|}{59.24} \\

        \multicolumn{1}{c|}{}&{\scriptsize FLeW-rs} & \textbf{68.82} & \textbf{85.25} & 72.65 & \textbf{75.57} & 46.59 & 59.96 & 84.52 & \textbf{43.62} & 79.83 & 87.48 & \multicolumn{1}{c|}{67.00} & \multicolumn{1}{c|}{58.30} \\
        \cmidrule(l){1-14}
        \multicolumn{2}{c}{{\scriptsize FLeW}} & 72.64 & 89.14 & \textbf{74.13} & \textbf{78.64} & 50.25 & \textbf{65.41} & \textbf{87.17} & \textbf{44.61} & \textbf{81.71} & 89.39 & \multicolumn{1}{c|}{\textbf{69.76}} & \multicolumn{1}{c|}{\textbf{60.81}} \\ 

        \multicolumn{2}{c}{\it{- w/o Textual}} & \textbf{72.89} & 88.79 & 74.00 & 78.56 & 49.81 & 64.47 & 86.83 & 44.55 & 81.57 & \textbf{89.83} & \multicolumn{1}{c|}{69.51} & \multicolumn{1}{c|}{60.66} \\

        \multicolumn{2}{c}{\it{- w/o Structural}} & 72.03 & \textbf{89.46} & 73.76 & 78.42 & \textbf{50.53} & 62.96 & 86.76 & 42.16 & 81.03 & 89.00 & \multicolumn{1}{c|}{68.74} & \multicolumn{1}{c|}{60.21} \\
        \midrule
        \end{tabular}
    }

\end{table*}

\section{Conclusion}
In this work, we propose FLeW to learn facet-level and adaptive weighted representations of scientific documents. Our approach unifies three key strategies for improved representation learning: citation-structural contrastive training, fine-grained multi-vector representation, and extra task-aware learning. We propose Structural Sampling, which leverages citation intent and frequency information to enhance citation-structural signals for contrastive training. By aligning citation intent with a generalized facet partition of scientific documents, we further propose Textual Splitting to split full text of triplets into faceted parts for better fine-grained textual embedding. Three facet encoders are trained on facet-specific triplets enriched with structural and textual information. With three encoders for representation, FLeW adopts a simple weight search to use the weighted sum of three facet representations as adaptive document representation instead of task-aware fine-tuning. Experiments show the applicability and robustness of FLeW across multiple scientific tasks and fields, compared to prior methods.

\begin{credits}
    \subsubsection{\ackname} This research work was supported by the National Natural Science Foundation of China (Grant No. 62276015).
\end{credits}
\bibliographystyle{splncs04}

\bibliography{FLeW}
\end{document}